\documentclass[aps,prl,twocolumn, superscriptaddress,groupedaddress]{revtex4}  

\usepackage{graphicx}
\usepackage{epsfig}
\usepackage{bm}
\usepackage{amsmath}
\usepackage{amssymb}
\usepackage{wasysym}
\usepackage{epstopdf}
\usepackage{color}
\usepackage{xcolor}
\usepackage{epstopdf}
\usepackage{upgreek}
\usepackage{natbib}
\usepackage{arydshln}
\usepackage[T1]{fontenc}
 
\usepackage{natbib}

\usepackage{psfrag}
\usepackage{color}
\usepackage{graphicx,float,amssymb}

\usepackage{color}
\usepackage{graphicx}
\usepackage{natbib}
\usepackage{bm}
\usepackage{amsmath}

\usepackage{graphicx}
\usepackage{epsfig}
\usepackage{bm}
\usepackage{amsmath}
\usepackage{amssymb}
\usepackage{wasysym}
\usepackage{epstopdf}
\usepackage{color}
\usepackage{xcolor}
\usepackage{epstopdf}
\usepackage{upgreek}
\usepackage{natbib}
\usepackage{arydshln}
\usepackage[T1]{fontenc}

\begin{document}

\title{Controlling heat transport and flow structures in thermal turbulence using ratchet surfaces}

\author{Hechuan Jiang}
\affiliation{Center for Combustion Energy and Department of Thermal Engineering, 
Tsinghua University, 100084 Beijing, China.}

\author{Xiaojue Zhu}
\affiliation{Physics of Fluids Group and Max Planck Center for Complex Fluid Dynamics, MESA+ Institute and J. M. Burgers Centre for Fluid Dynamics, University of Twente, P.O. Box 217, 7500AE Enschede, The Netherlands}

\author{Varghese Mathai}
\affiliation{Physics of Fluids Group and Max Planck Center for Complex Fluid Dynamics, MESA+ Institute and J. M. Burgers Centre for Fluid Dynamics, University of Twente, P.O. Box 217, 7500AE Enschede, The Netherlands}

\author{Roberto Verzicco}
\affiliation{Dipartimento di Ingegneria Industriale, University of Rome "Tor Vergata", Via del Politecnico 1, Roma 00133, Italy}
\affiliation{Physics of Fluids Group and Max Planck Center for Complex Fluid Dynamics, MESA+ Institute and J. M. Burgers Centre for Fluid Dynamics, University of Twente, P.O. Box 217, 7500AE Enschede, The Netherlands}

\author{Detlef Lohse}
\affiliation{Physics of Fluids Group and Max Planck Center for Complex Fluid Dynamics, MESA+ Institute and J. M. Burgers Centre for Fluid Dynamics, University of Twente, P.O. Box 217, 7500AE Enschede, The Netherlands}
\affiliation{
Max Planck Institute for Dynamics and Self-Organization, 37077 G\"ottingen, Germany.}
\affiliation{Center for Combustion Energy and Department of Thermal Engineering, 
Tsinghua University, 100084 Beijing, China.}

\author{Chao Sun}
\thanks{chaosun@tsinghua.edu.cn}
\affiliation{Center for Combustion Energy and Department of Thermal Engineering, 
Tsinghua University, 100084 Beijing, China.}
\affiliation{Physics of Fluids Group and Max Planck Center for Complex Fluid Dynamics, MESA+ Institute and J. M. Burgers Centre for Fluid Dynamics, University of Twente, P.O. Box 217, 7500AE Enschede, The Netherlands}

\date{\today}

\date{\today}

\begin{abstract} 

In this combined experimental and numerical study on thermally driven turbulence in a rectangular cell, the global heat transport and the coherent flow structures are controlled with an asymmetric ratchet-like roughness on the top and bottom plates. We show that, by means of symmetry breaking  due to the presence of the ratchet structures on the conducting plates, the orientation of the Large Scale Circulation Roll (LSCR) can be locked to a preferred direction even when the cell is perfectly leveled out. By introducing a small tilt to the system, we show that the LSCR orientation can be tuned and controlled. The two different orientations of LSCR give two quite different heat transport efficiencies, indicating that heat transport is sensitive to the LSCR direction over the asymmetric roughness structure. Through a quantitative analysis of the dynamics of thermal plume emissions and the orientation of the LSCR over the asymmetric structure, we provide a physical explanation for these findings. The current work has important implications for passive and active flow control in engineering, bio-fluid dynamics, and geophysical flows.
\end{abstract}

\maketitle

Turbulent convective flows over rough surfaces are ubiquitous in  engineering and geophysical flows. Examples include convective flows in the atmosphere and in oceans, where the ground,  sea-bed and ocean floor are generally not smooth. As the ability to enhance convective heat transfer is crucial in many industrial applications, numerous strategies have been proposed to efficiently enhance it. Among several approaches, introducing wall roughness is an effective way  to do so.  
Indeed, the study of surface roughness effects in wall-bounded turbulent flows has been an area of intense research~(see e.g. some recent work \cite{nik33,ber03,sho06,hul13,cha15,chu15,squ16,mac16}, the reviews \cite{jim04,fla14}, and the textbooks \cite{pop00,sch00}). Similarly, several studies have been conducted on turbulent thermal convection over rough plates~\cite{she96,du00,roc01,qiu05,tis11,sal14,wag15,top17}.
The vast majority of these studies with rough walls adopt some ordered and symmetrical structures, such as pyramids, squares, rectangles etc. However, the rough surfaces in engineering applications and in nature are in general not symmetric, resulting in complex interactions between the flow and the asymmetric roughness elements.
Examples are wind blowing over a landscape with asymmetric slopes and ocean flows over an asymmetric sea-bed, etc. Other examples include marine animals which can actively change the asymmetric roughness for maneuverability.

In this work, we aim to study the influences of ratchet-like wall structures on the flow organization and heat transfer in fully developed convective thermal turbulence. 
Indeed, building on the classical Feynman-Smoluchowski ratchet, in various contexts researchers have proposed ratchet-type mechanisms and devices that operate outside of thermal equilibrium \cite{rei02}. Examples include the so-called `capillary-ratchet' in zoology \cite{pra08}, a rotational ratchet in a granular
gas \cite{esh10}, self-propelled Leidenfrost droplets and solids on ratchet surfaces \cite{lin06,lag11,marin2012capillary}, and the Brownian ratchets of molecular motors in living organisms \cite{han09}. 
Here we will employ the classical model system for thermal turbulence, namely Rayleigh-B\'enard (RB) convection \cite{ahl09,loh10}, to investigate the effects of asymmetric roughness on convective heat transfer and large scale flow organization. RB convection consists of a working fluid confined between a cold top (T$_t$) and a warm bottom plate (T$_{b}$), with a constant temperature difference ($\Delta$ = T$_{b} -$ T$_t$). The dynamics of the system depends on the driving intensity and the fluid properties, which are characterized, respectively, only by the Rayleigh number $Ra=\frac{\alpha g \text{H}^3\Delta}{\nu\kappa}$ and the Prandtl number $Pr=\frac{\nu}{\kappa}$, 
where $\alpha$, $g$, H, $\nu$, and $\kappa$ are the thermal expansion coefficient, the acceleration due to gravity, the thickness of the fluid layer, kinematic viscosity, and thermal diffusivity of the convecting fluid, respectively. 
The key response parameter of the RB system is the non-dimensional heat flux, the Nusselt number $Nu={J}/{(\chi \ \Delta/\text{H})}$, which measures the ratio of the heat flux ($J$) over the purely conductive (thermal conductivity $\chi$) one.

\begin{figure*}
\centering
\includegraphics[width=1\linewidth]{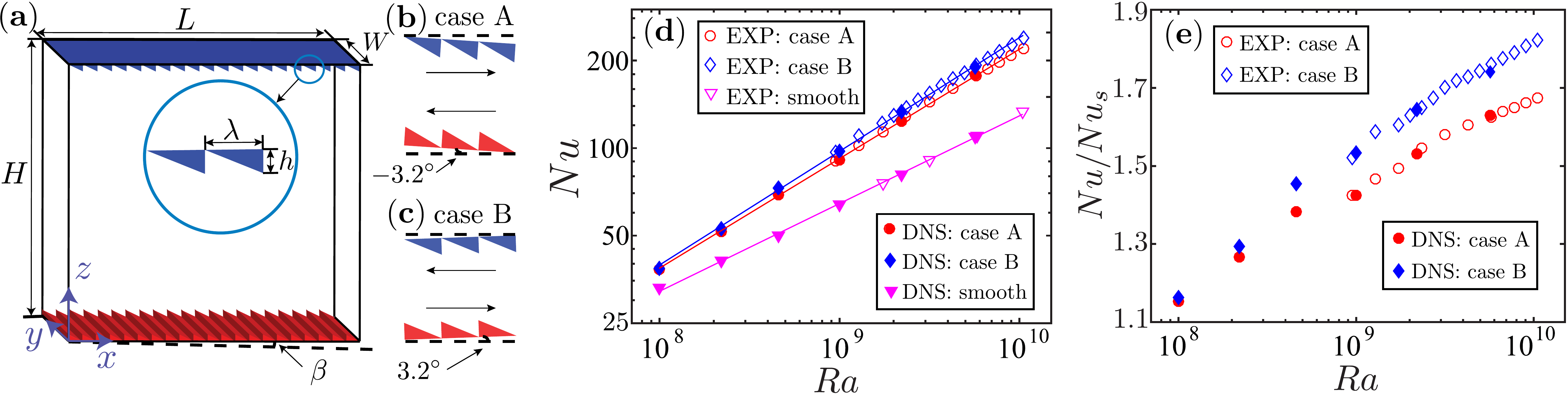}
\caption{(a) A sketch of the convection cell with the asymmetric ratchets on the top and bottom plates. The cell can be titled to the left or the right side.  (b) Tilting angle $\beta = -3.2^\circ$ resulting in a flow in clockwise direction. (c) Tilting angle $\beta = + 3.2^\circ$ resulting in a flow in counter-clockwise direction. (d) Nusselt number $Nu$ as a function of $Ra$ in the smooth and rough cells ($Pr$ = 4.3). (e) $Nu$ enhancement as a function of $Ra$ for the two cases. The open symbols correspond to experimental data, the closed symbols correspond to numerical data, and the triangles represent the data for the smooth wall case.}
\label{fig:setup}
\end{figure*}

In our experiments, a novel thermal convection system (sketched in Fig.~\ref{fig:setup}) with ratchet structures on the top and bottom plates is used.  The convection cell is of rectangular shape, with upper and lower plates made of copper, and plexiglas sidewalls. The length (L), width (W), and height (H) are 240 mm, 60 mm, and 240 mm, respectively, resulting in a unit aspect ratio in the large scale circulation plane ($\Gamma = \text{W/H}$). The bottom plate is heated at a constant heat flux, and the top plate temperature is regulated at constant temperature. The ratchet-like structures are machined on the lower and upper surfaces, respectively, of the top and bottom plates of the convection cell. The height h and the width $\lambda$ of the ratchets are 6 mm and 12 mm, respectively. Six thermistors (Omega 44131) are embedded in the top and bottom plates to probe the local temperature in the plates. To control the orientation of flow in the main LSCR plane, the cell can be tilted  in both clockwise and counter-clockwise directions with some angle $\beta$. 

In addition to the experiments, three-dimensional direct numberical simulations~(DNS) of the Boussinesq equations are performed by using the in-house AFiD code \cite{poe15cf,zhu17afid}. An immersed boundary method~\cite{fadlun00} is implemented to simulate the ratchet surfaces. The code has been extensively validated and used \cite{poe15cf,zhu17afid,zhu17b}. Adequate resolutions~\cite{supplemental} are ensured for all simulations so that the results are grid independent. At $Ra=5.7\times10^9$, $1280\times1280\times256$ grid points are used for the cases with ratchet. The grid used is fine near the boundaries, and gradually grows toward the bulk region. This results in about 24 grid points within the boundary layer height, which is sufficient to capture the boundary layers~\cite{shi10}. Similar to the experiments, the no-slip boundary conditions are adopted for the velocity at all solid boundaries. At all side walls the heat-insulating conditions are adopted, and at top and bottom plates constant temperatures are prescribed. At high $Ra$, the difference in global heat transfer between constant temperature and constant heat flux boundary conditions is known to be small~\cite{johnston2009comparison,huang2015comparative}. Therefore, it is reasonable to compare $Nu$ from experiments and simulations.

We first study how the orientation of the LSCR affects the convective heat transfer over the ratchets. In order to prescribe the orientation of the LSCR, we tilt the convection cell by either $+3.2^\circ$ or $-3.2^\circ$. This little tilting hardly affects Nu \cite{shi16}; here we find an effect of less than 2\% in all cases, much less than through roughness or the orientation of the roughness. 
The aim of the tilting is to lock the LSCR direction as sketched in Figs.~\ref{fig:setup}(b,c). In clockwise direction (tilting angle of $-$3.2$^\circ$), the flow near the top and bottom plates moves along the smaller slope side of the ratchets. We refer to this situation as case A, whereas the situation where the flow near the top and bottom plates travels toward the steeper side of the ratchets is referred to as case B (see Fig.~\ref{fig:setup}(c) where the tilting angle is +3.2$^\circ$). 

\begin{figure*} [!htbp]
\centering
\includegraphics[width=1.0\linewidth]{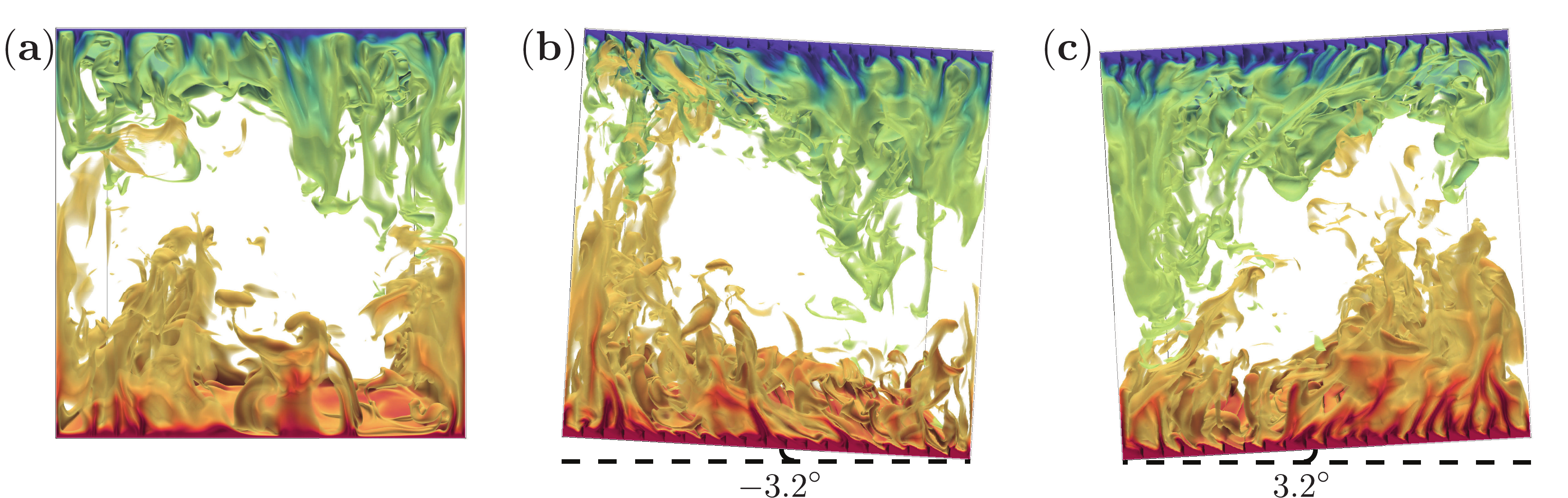}
\caption{Volume rendering of the temperature field, showing plume dynamics at $Ra$ = 5.7 $\times 10^9$ and Pr = 4.3 from the DNS for (a) the smooth wall situation; (b)  case A; (c)  case B. The corresponding movies are available in the Supplemental Material.}
\label{fig:dnsimages}
\end{figure*}
The Prandtl number is fixed at $Pr$ = 4.3 for all of the measurements.  Fig.~\ref{fig:setup}(d) shows the measured $Nu$ as a function of $Ra$ for the cases A \& B. For comparison, we also measure the data at the same RB system with a smooth top and bottom plate. 
$Nu$($Ra$) can be described with a power law with an effective exponent of 0.30 $\pm$ 0.01. For the ratchet surfaces, it is found that $Nu$($Ra$) are much larger than that for the smooth wall case and have steeper effective slopes with 0.38 $\pm$ 0.01 for case A and 0.39 $\pm$ 0.01 for case B. The higher effective exponent observed for the rough cell as compared to that of the smooth cell has been explained in a recent work~\cite{zhu17b}. 
The simulated $Nu$ as a function of $Ra$ is also shown in Fig.~\ref{fig:setup}(d). We find excellent agreement between the experiments and numerics. In Fig.~\ref{fig:setup}(e), we plot the $Nu$ enhancement by dividing the data in the rough cell by $Nu$ of the smooth cell. It is clearly seen that in both cases A \& B, $Nu$ is enhanced, and interestingly, this enhancement increases with $Ra$.
The latter trend can be explained as follows: The  thermal boundary layer thickness $\delta_T$ decreases with increasing $Ra$, leading to an increase in the effective roughness height $\text{h}/\delta_T$; Correspondingly, the roughness elements penetrate more deeply into the thermal boundary layers,  thereby triggering stronger plume emissions, which explains the greater $Nu$ enhancement. \textcolor{black}{We find that $Nu$ for the symmetric case is almost the same as that for case A \cite{supplemental}.}

We now come to the main subject of this paper, and compare and contrast the $Nu$ enhancement between the two cases of different roughness orientation with respect to the LSCR (cases A \& B). At the lowest Ra ($\sim 10^8$), the $Nu$ enhancements are relatively small and almost indistinguishable for the two cases~(15.4\% for case A, and 16.3\% for case B). In this situation, the thermal boundary layer thickness $\delta_T \sim 4$ mm is comparable to the roughness height h = 6 mm  \cite{qiu05}.
Thus, the orientation of the flow has negligible influence on $Nu$.  However, the $Nu$ enhancements for the two cases become increasingly different at larger $Ra$. At the largest $Ra \approx 10^{10}$, $Nu/Nu_s = $ 67.4\% for case A, as against 82.2\% for case B. 
What is the physical reason for this significant difference in $Nu$ enhancement in the two cases?

Fig.~\ref{fig:dnsimages} shows three instantaneous temperature fields at $Ra$ = 5.7 $\times 10^9$ for the smooth wall case,  case A, and case B. In the smooth wall case, the spots where plumes detach from the boundary layers to the bulk are randomly located near the plates, as shown in Fig.~\ref{fig:dnsimages}(a). For case A, the flow moves on the ratchet structures, but faces the ratchet side with the smaller slope.  The horizontal motion of the flow near the wall are modified by the presence of the roughness, and consequently the plumes are detached from some of the ratchet tips, as shown in Fig.~\ref{fig:dnsimages}(b). Thanks to these sharp tips, more plumes are ejected from the boundary layers to the bulk, resulting in a higher heat transfer than the smooth plate case. Nevertheless, the overall features of the flow near the boundary layers are quite similar to those of the smooth plate case. The situation becomes very different in case B shown in Fig.~\ref{fig:dnsimages}(c). The horizontal LSCR now hits the sharp corners of the ratchets, resulting in  strong plume detachments at the many tips of the structures as seen in Fig.~\ref{fig:dnsimages}(c).  Clearly, more plumes are emitted from the boundary layers to the bulk, explaining the even higher $Nu$ than that of case A.

\begin{figure}[!hbp]
	\centering
	\includegraphics[width=1.0\linewidth]{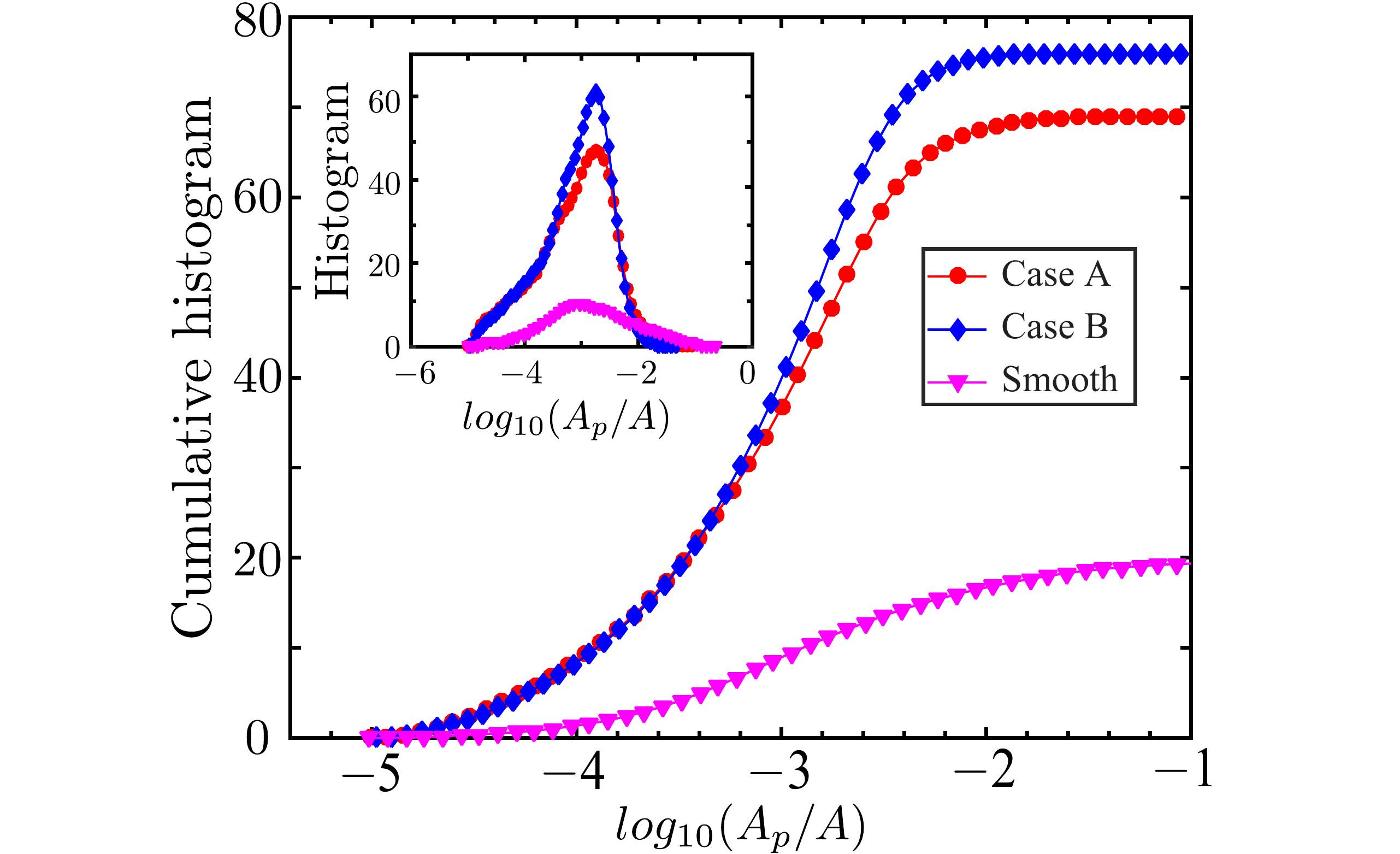}
	\caption{Cumulative histogram of the normalized plume area $A_p/A$ for  case A, and case B, smooth case. Inset shows the  histograms of the same. Case B shows the highest number of plume emissions.}
	\label{fig:plume_cdf_and_pdf}
\end{figure}

\begin{figure}[!tbp]
\centering
\includegraphics[width=1.0\linewidth]{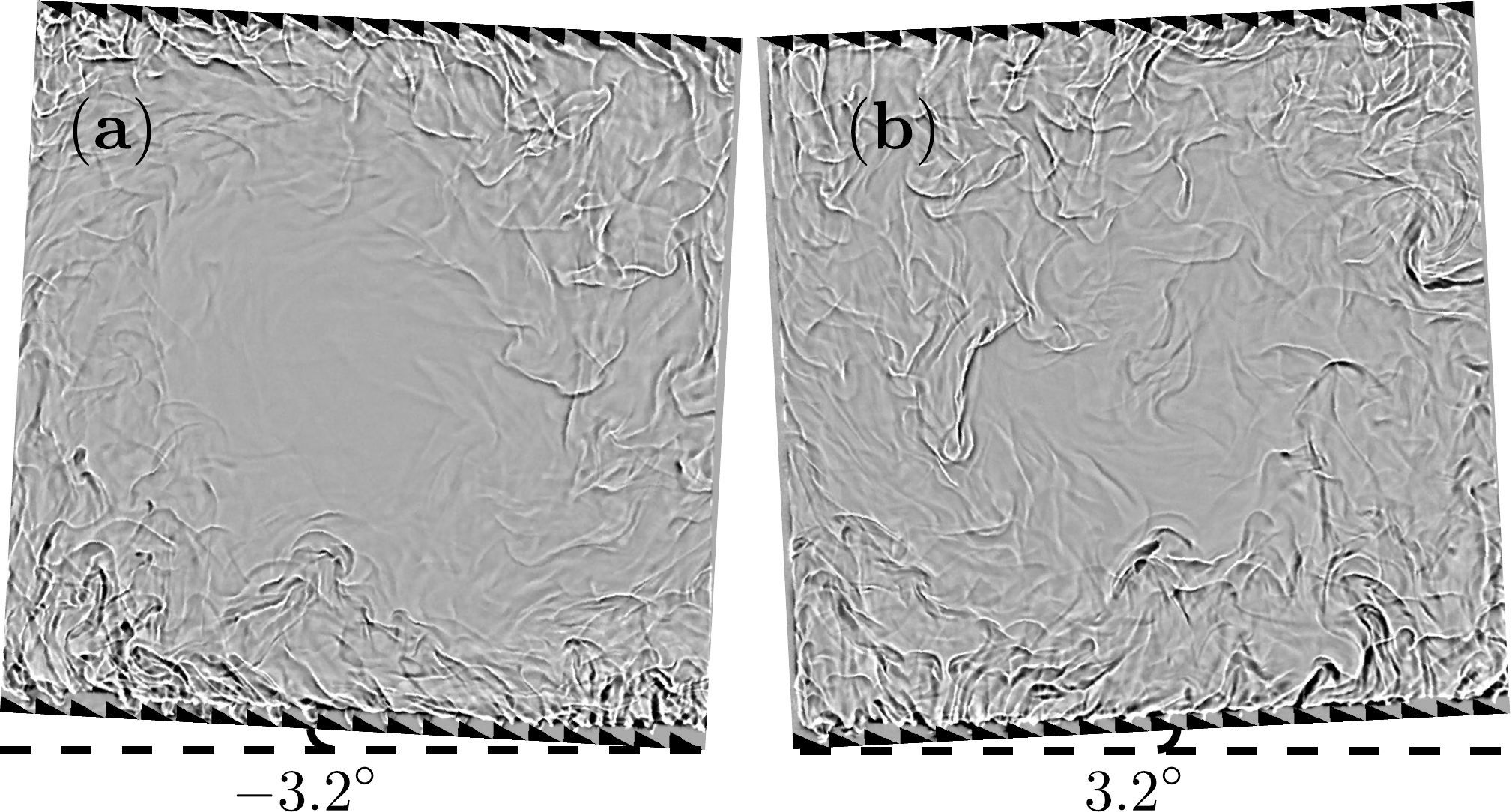}
\caption{Shadowgraph visualization  of the spatial distribution of thermal plumes at $Ra$ = 5.7 $\times 10^9$ and $Pr$ = 5.7 for (a) case A, and (b) case B. \textcolor{black}{The Large Scale Circulation Roll direction is clockwise in (a), and counter-clockwise in (b).} The corresponding movies are available in the Supplemental material.}
\label{fig:shadowgraph}
\end{figure}

\textcolor{black}{Next, we quantify the plume emissions for the three cases, employing the method introduced in Refs.~\cite{huang13,poe15jfm} (see the Supplemental Material~\cite{supplemental} for details).
Fig.~\ref{fig:plume_cdf_and_pdf} shows the cumulative histogram of the plume areas $A_p$ normalized by the plate area $A$ at a height $z/H = 0.028$. Case B has the highest number of plumes, followed by case A, and then, the smooth case. The inset shows the histogram of the plume areas, which indicates that most of the plumes emitted for cases A and B have an intermediate area, with the number of plumes in case B exceeding that in case A.} Further, we estimated the velocity of the LSCR for case A and case B. Interestingly, case B has a larger roll velocity, $V_{\text{LSC}}$(B) = 0.129, than case A, for which $V_{\text{LSC}}$(A) = 0.117, reflecting that more plumes lead to a  stronger LSC. This is due to plume driving, which finally leads to a higher heat transfer~\cite{xi04}.
Figs. \ref{fig:shadowgraph}(a,b) show the shadow-graphic visualizations at these two different LSCR orientations \cite{xi04}. Due to the small size of the plumes, it is difficult to appreciate the differences between case A and B in the wall region. However the bulk of case B is clearly more plume dominated indicating the more plumes are released from the walls. Thanks to the much stronger detachment of thermal boundary layer in case B than that in case A, the temperature distribution of the bottom plate is relatively uniform. A detailed discussion is given in the Supplemental Material~\cite{supplemental}.

\begin{figure} [!hp]
\centering
\includegraphics[width=1.0\linewidth]{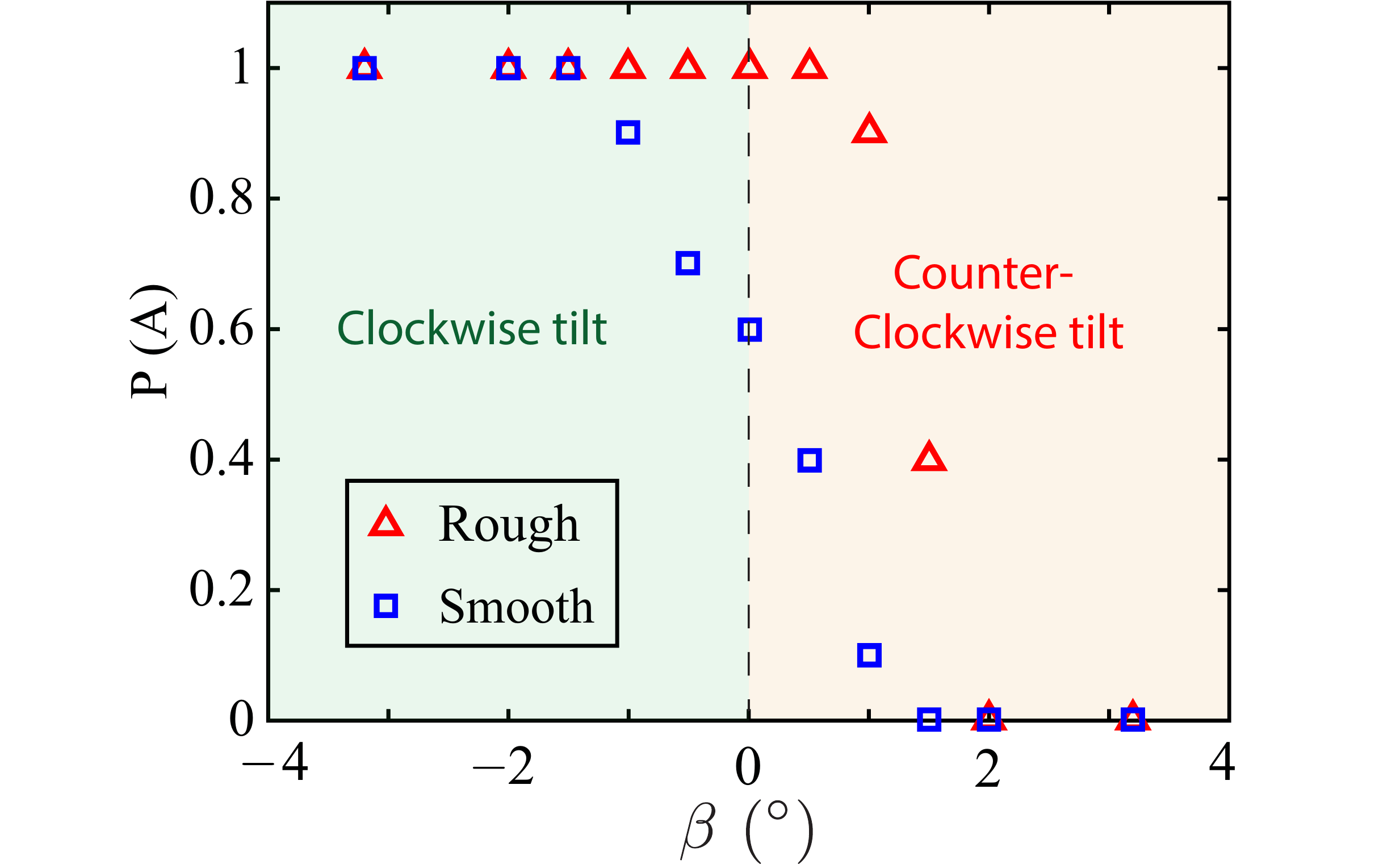}
\caption{The measured probability for the LSCR orientation to remain in case A as a function of tilting angle at $Ra$ = 5.7 $\times 10^9$. A value of 1 means that the LSCR orientation is always in case A, whereas a value 0 means that it is in case B. 
In the transition regime, each measurements are repeated 10 times. 
}
\label{fig:lscor}
\end{figure}

Finally, we study how the system decides the LSCR orientation for various tilting angles $\beta$. 
We determine the direction of the LSCR by simply measuring maximum temperature difference in the bottom plate, as discussed above, and as done previously in Refs. \cite{sui10,ni15,hua15}.
First, for $\beta = 0^\circ$, when the system with ratchets is leveled, the LSCR orientation is always as in case A, in which the viscous drag induced by the rough elements is smaller as compared to that of the case B.
Of course, the flow is in the same situation when the system is tilted at the negative angles, and it is locked in this case A even if the system is titled at a positive angle up to 0.5 degree. To enforce a counter-clockwise LSCR throughout (case B), one has to tilt the system by at least 2 degree. In between 0.5 and 2 degrees, the system can be in either of the two states. We measure the flow states in multiple experiments and determine the probability of a certain LSCR orientation as a function of the tilting angle $\beta$. Fig.~\ref{fig:lscor} plots the probability of the LSCR in case A, which is defined as $P(A) =$ (\# in case A)$/$(Total \# of the measurements). For comparison, we also perform measurements in the smooth wall system. 

The transition range of the tilting angle from full case A state (P(A) = 1) to full case B state (P(A) = 0) is from $\beta \approx -1.5^\circ$ to $\approx 1.5^\circ$ for the smooth wall case, whereas the transition range for the ratchet wall system is much sharper, i.e. from $\beta \approx 0.5 ^\circ$ to $\approx 2^\circ$, indicating that the LSCR is more sensitive to the tilt in the ratchet wall system. 
As expected, for the smooth wall case, the measured P(A) shows a symmetric trend with respect to $\beta = 0^\circ$, i.e. P(A) $\approx$ 0.5 at $\beta_c$ = 0$^\circ$;  With the current ratchet plates, the symmetric center shifts to $\beta_c \approx 1^\circ-1.5^\circ$.

The asymmetry observed in $P(A)$ of the LSCR (Fig.~\ref{fig:lscor}) can be rationalized by considering the asymmetry of the roughness elements. We observe $P(A)  = 1$ for $\beta \leq 0$, since the case A is strongly favored due to the asymmetric shape of the ratchets. However, when $\beta$ is increased above $0^\circ$, a component of the buoyancy force $F_{b\beta} \approx \frac{1}{2} \rho \mathcal{V} g \ \alpha \Delta  \sin \beta$ favors the case B flow. As we keep on increasing $\beta$, at a certain critical angle $\beta_c$, this buoyancy force might overcome the drag force opposing the case B flow. For the finite size of the rougness elements in our case, we expect the pressure contribution to be dominant~\cite{zhu17}, so that the drag force may be written as: $F_d \approx \frac{1}{2}\rho v_f^2 A_p n \ C_{D,B} $. Here, $v_f \approx 0.2 \ \frac{\nu}{H} \sqrt {\frac{Ra}{Pr}}$ is the flow velocity seen by the ratchets, $\mathcal{V}$ is the effective volume of the heated/cooled fluid, $A_p$ is the projected area of a ratchet in the plane that is perpendicular to the flow direction, $C_{D,B} \approx 2$ is the drag coefficient for flow past the ratchet structure, which is modeled as a triangular half-body facing a flow, and $n$ is the number of the ratchets on the plate.  Equating the buoyancy and drag forces, and plugging in the numbers at this Ra ($\Delta$ = 10.9 K, $\nu$ = 6.6 $\times$ 10$^{-7}$ m$^2/s$, H = 0.24 m, $\rho$ = 992.2 kg/m$^3$, n = 20, $\mathcal{V}$ =3.5 $\times$ 10$^{-3}$ m$^3$, $\alpha$ = 3.9 $\times$ 10$^{-4}$ K$^{-1}$), we obtain a critical angle estimate $\beta_c \approx 2^{\circ}$, which compares fairly well with our experimental observation.
This symmetrical center may have some dependence on $Ra$ and $Pr$, which deserves future studies. 
In the system with the ratchet walls, we do not see a single reversal event in all of our measurements (410 hours in total, around 30,000 turn over time), indicating that the LSCR orientation in the ratchet cell is very stable, but meanwhile can be controlled.

In summary, we find that the global heat transport is sensitive to the LSCR direction over asymmetric surface structure, and provide a physical understanding for the heat transport difference. The LSCR orientation has a preferred direction when the cell is perfectly leveled, but it can be tuned and controlled by introducing a marginal tilt to the system. 
This provides many flow control opportunities to achieve stable flow structures which avoid unpredictable flow reversal events, and to homogenize the wall temperature distribution in complex flow environments. Further, active switching of the ratchet type boundary condition in one or the other direction could offer a clean way of controlling the flow  and heat transfer. 

\begin{acknowledgments}
We thank the anonymous reviewers for constructive suggestions on the manuscript. We gratefully acknowledge Linfeng Jiang for his help with experiments. 
 This work is financially supported by the Natural Science Foundation of China under Grant No. 11672156, and the Dutch Foundation for
Fundamental Research on Matter (NWO).
\end{acknowledgments}


\begin{thebibliography}{45}
	\expandafter\ifx\csname natexlab\endcsname\relax\def\natexlab#1{#1}\fi
	\expandafter\ifx\csname bibnamefont\endcsname\relax
	\def\bibnamefont#1{#1}\fi
	\expandafter\ifx\csname bibfnamefont\endcsname\relax
	\def\bibfnamefont#1{#1}\fi
	\expandafter\ifx\csname citenamefont\endcsname\relax
	\def\citenamefont#1{#1}\fi
	\expandafter\ifx\csname url\endcsname\relax
	\def\url#1{\texttt{#1}}\fi
	\expandafter\ifx\csname urlprefix\endcsname\relax\def\urlprefix{URL }\fi
	\providecommand{\bibinfo}[2]{#2}
	\providecommand{\eprint}[2][]{\url{#2}}
	
	\bibitem[{\citenamefont{Nikuradse}(1933)}]{nik33}
	\bibinfo{author}{\bibfnamefont{J.}~\bibnamefont{Nikuradse}},
	\bibinfo{journal}{Forschungsheft Arb. Ing.-Wes.}
	\textbf{\bibinfo{volume}{361}} (\bibinfo{year}{1933}).
	
	\bibitem[{\citenamefont{van~den Berg et~al.}(2003)\citenamefont{van~den Berg,
			Doering, Lohse, and Lathrop}}]{ber03}
	\bibinfo{author}{\bibfnamefont{T.~H.} \bibnamefont{van~den Berg}},
	\bibinfo{author}{\bibfnamefont{C.}~\bibnamefont{Doering}},
	\bibinfo{author}{\bibfnamefont{D.}~\bibnamefont{Lohse}}, \bibnamefont{and}
	\bibinfo{author}{\bibfnamefont{D.}~\bibnamefont{Lathrop}},
	\bibinfo{journal}{Phys. Rev. E} \textbf{\bibinfo{volume}{68}},
	\bibinfo{pages}{036307} (\bibinfo{year}{2003}).
	
	\bibitem[{\citenamefont{Shockling et~al.}(2006)\citenamefont{Shockling, Allen,
			and Smits}}]{sho06}
	\bibinfo{author}{\bibfnamefont{M.~A.} \bibnamefont{Shockling}},
	\bibinfo{author}{\bibfnamefont{J.~J.} \bibnamefont{Allen}}, \bibnamefont{and}
	\bibinfo{author}{\bibfnamefont{A.~J.} \bibnamefont{Smits}},
	\bibinfo{journal}{J. Fluid Mech.} \textbf{\bibinfo{volume}{564}},
	\bibinfo{pages}{267} (\bibinfo{year}{2006}).
	
	\bibitem[{\citenamefont{Hultmark et~al.}(2013)\citenamefont{Hultmark,
			Vallikivi, Bailey, and Smits}}]{hul13}
	\bibinfo{author}{\bibfnamefont{M.}~\bibnamefont{Hultmark}},
	\bibinfo{author}{\bibfnamefont{M.}~\bibnamefont{Vallikivi}},
	\bibinfo{author}{\bibfnamefont{S.~C.~C.} \bibnamefont{Bailey}},
	\bibnamefont{and} \bibinfo{author}{\bibfnamefont{A.~J.} \bibnamefont{Smits}},
	\bibinfo{journal}{J. Fluid Mech.} \textbf{\bibinfo{volume}{728}},
	\bibinfo{pages}{376} (\bibinfo{year}{2013}).
	
	\bibitem[{\citenamefont{Chan et~al.}(2015)\citenamefont{Chan, MacDonald, Chung,
			Hutchins, and Ooi}}]{cha15}
	\bibinfo{author}{\bibfnamefont{L.}~\bibnamefont{Chan}},
	\bibinfo{author}{\bibfnamefont{M.}~\bibnamefont{MacDonald}},
	\bibinfo{author}{\bibfnamefont{D.}~\bibnamefont{Chung}},
	\bibinfo{author}{\bibfnamefont{N.}~\bibnamefont{Hutchins}}, \bibnamefont{and}
	\bibinfo{author}{\bibfnamefont{A.}~\bibnamefont{Ooi}}, \bibinfo{journal}{J.
		Fluid Mech.} \textbf{\bibinfo{volume}{771}}, \bibinfo{pages}{743}
	(\bibinfo{year}{2015}).
	
	\bibitem[{\citenamefont{Chung et~al.}(2015)\citenamefont{Chung, Chan,
			MacDonald, Hutchins, and Ooi}}]{chu15}
	\bibinfo{author}{\bibfnamefont{D.}~\bibnamefont{Chung}},
	\bibinfo{author}{\bibfnamefont{L.}~\bibnamefont{Chan}},
	\bibinfo{author}{\bibfnamefont{M.}~\bibnamefont{MacDonald}},
	\bibinfo{author}{\bibfnamefont{N.}~\bibnamefont{Hutchins}}, \bibnamefont{and}
	\bibinfo{author}{\bibfnamefont{A.}~\bibnamefont{Ooi}}, \bibinfo{journal}{J.
		Fluid Mech.} \textbf{\bibinfo{volume}{773}}, \bibinfo{pages}{418}
	(\bibinfo{year}{2015}).
	
	\bibitem[{\citenamefont{Squire et~al.}(2016)\citenamefont{Squire,
			Morrill-Winter, Hutchins, Schultz, Klewicki, and Marusic}}]{squ16}
	\bibinfo{author}{\bibfnamefont{D.~T.} \bibnamefont{Squire}},
	\bibinfo{author}{\bibfnamefont{C.}~\bibnamefont{Morrill-Winter}},
	\bibinfo{author}{\bibfnamefont{N.}~\bibnamefont{Hutchins}},
	\bibinfo{author}{\bibfnamefont{M.~P.} \bibnamefont{Schultz}},
	\bibinfo{author}{\bibfnamefont{J.~C.} \bibnamefont{Klewicki}},
	\bibnamefont{and} \bibinfo{author}{\bibfnamefont{I.}~\bibnamefont{Marusic}},
	\bibinfo{journal}{J. Fluid Mech.} \textbf{\bibinfo{volume}{795}},
	\bibinfo{pages}{210} (\bibinfo{year}{2016}).
	
	\bibitem[{\citenamefont{MacDonald et~al.}(2016)\citenamefont{MacDonald, Chan,
			Chung, Hutchins, and Ooi}}]{mac16}
	\bibinfo{author}{\bibfnamefont{M.}~\bibnamefont{MacDonald}},
	\bibinfo{author}{\bibfnamefont{L.}~\bibnamefont{Chan}},
	\bibinfo{author}{\bibfnamefont{D.}~\bibnamefont{Chung}},
	\bibinfo{author}{\bibfnamefont{N.}~\bibnamefont{Hutchins}}, \bibnamefont{and}
	\bibinfo{author}{\bibfnamefont{A.}~\bibnamefont{Ooi}}, \bibinfo{journal}{J.
		Fluid Mech.} \textbf{\bibinfo{volume}{804}}, \bibinfo{pages}{130}
	(\bibinfo{year}{2016}).
	
	\bibitem[{\citenamefont{Jim\'enez}(2004)}]{jim04}
	\bibinfo{author}{\bibfnamefont{J.}~\bibnamefont{Jim\'enez}},
	\bibinfo{journal}{Ann. Rev. Fluid Mech.} \textbf{\bibinfo{volume}{36}},
	\bibinfo{pages}{173} (\bibinfo{year}{2004}).
	
	\bibitem[{\citenamefont{Flack and Schultz}(2014)}]{fla14}
	\bibinfo{author}{\bibfnamefont{K.~A.} \bibnamefont{Flack}} \bibnamefont{and}
	\bibinfo{author}{\bibfnamefont{M.~P.} \bibnamefont{Schultz}},
	\bibinfo{journal}{Phys. Fluids} \textbf{\bibinfo{volume}{26}},
	\bibinfo{pages}{101305} (\bibinfo{year}{2014}).
	
	\bibitem[{\citenamefont{Pope}(2000)}]{pop00}
	\bibinfo{author}{\bibfnamefont{S.~B.} \bibnamefont{Pope}},
	\emph{\bibinfo{title}{Turbulent Flow}} (\bibinfo{publisher}{Cambridge
		University Press}, \bibinfo{address}{Cambridge}, \bibinfo{year}{2000}).
	
	\bibitem[{\citenamefont{Schlichting and Gersten}(2000)}]{sch00}
	\bibinfo{author}{\bibfnamefont{H.}~\bibnamefont{Schlichting}} \bibnamefont{and}
	\bibinfo{author}{\bibfnamefont{K.}~\bibnamefont{Gersten}},
	\emph{\bibinfo{title}{Boundary layer theory}} (\bibinfo{publisher}{Springer
		Verlag}, \bibinfo{address}{Berlin}, \bibinfo{year}{2000}),
	\bibinfo{edition}{8th} ed.
	
	\bibitem[{\citenamefont{Shen et~al.}(1996)\citenamefont{Shen, Tong, and
			Xia}}]{she96}
	\bibinfo{author}{\bibfnamefont{Y.}~\bibnamefont{Shen}},
	\bibinfo{author}{\bibfnamefont{P.}~\bibnamefont{Tong}}, \bibnamefont{and}
	\bibinfo{author}{\bibfnamefont{K.-Q.} \bibnamefont{Xia}},
	\bibinfo{journal}{Phys. Rev. Lett.} \textbf{\bibinfo{volume}{76}},
	\bibinfo{pages}{908} (\bibinfo{year}{1996}).
	
	\bibitem[{\citenamefont{Du and Tong}(2000)}]{du00}
	\bibinfo{author}{\bibfnamefont{Y.~B.} \bibnamefont{Du}} \bibnamefont{and}
	\bibinfo{author}{\bibfnamefont{P.}~\bibnamefont{Tong}}, \bibinfo{journal}{J.
		Fluid Mech.} \textbf{\bibinfo{volume}{407}}, \bibinfo{pages}{57}
	(\bibinfo{year}{2000}).
	
	\bibitem[{\citenamefont{Roche et~al.}(2001)\citenamefont{Roche, Castaing,
			Chabaud, and Hebral}}]{roc01}
	\bibinfo{author}{\bibfnamefont{P.~E.} \bibnamefont{Roche}},
	\bibinfo{author}{\bibfnamefont{B.}~\bibnamefont{Castaing}},
	\bibinfo{author}{\bibfnamefont{B.}~\bibnamefont{Chabaud}}, \bibnamefont{and}
	\bibinfo{author}{\bibfnamefont{B.}~\bibnamefont{Hebral}},
	\bibinfo{journal}{Phys. Rev. E} \textbf{\bibinfo{volume}{63}},
	\bibinfo{pages}{045303} (\bibinfo{year}{2001}).
	
	\bibitem[{\citenamefont{Qiu et~al.}(2005)\citenamefont{Qiu, Xia, and
			Tong}}]{qiu05}
	\bibinfo{author}{\bibfnamefont{X.~L.} \bibnamefont{Qiu}},
	\bibinfo{author}{\bibfnamefont{K.-Q.} \bibnamefont{Xia}}, \bibnamefont{and}
	\bibinfo{author}{\bibfnamefont{P.}~\bibnamefont{Tong}}, \bibinfo{journal}{J.
		Turb.} \textbf{\bibinfo{volume}{6}}, \bibinfo{pages}{1}
	(\bibinfo{year}{2005}).
	
	\bibitem[{\citenamefont{Tisserand et~al.}({2011})\citenamefont{Tisserand,
			Creyssels, Gasteuil, Pabiou, Gibert, Castaing, and Chilla}}]{tis11}
	\bibinfo{author}{\bibfnamefont{J.~C.} \bibnamefont{Tisserand}},
	\bibinfo{author}{\bibfnamefont{M.}~\bibnamefont{Creyssels}},
	\bibinfo{author}{\bibfnamefont{Y.}~\bibnamefont{Gasteuil}},
	\bibinfo{author}{\bibfnamefont{H.}~\bibnamefont{Pabiou}},
	\bibinfo{author}{\bibfnamefont{M.}~\bibnamefont{Gibert}},
	\bibinfo{author}{\bibfnamefont{B.}~\bibnamefont{Castaing}}, \bibnamefont{and}
	\bibinfo{author}{\bibfnamefont{F.}~\bibnamefont{Chilla}},
	\bibinfo{journal}{{Phys. Fluids}} \textbf{\bibinfo{volume}{{23}}},
	\bibinfo{pages}{015105} (\bibinfo{year}{{2011}}).
	
	\bibitem[{\citenamefont{Salort et~al.}({2014})\citenamefont{Salort, Liot,
			Rusaouen, Seychelles, Tisserand, Creyssels, Castaing, and Chilla}}]{sal14}
	\bibinfo{author}{\bibfnamefont{J.}~\bibnamefont{Salort}},
	\bibinfo{author}{\bibfnamefont{O.}~\bibnamefont{Liot}},
	\bibinfo{author}{\bibfnamefont{E.}~\bibnamefont{Rusaouen}},
	\bibinfo{author}{\bibfnamefont{F.}~\bibnamefont{Seychelles}},
	\bibinfo{author}{\bibfnamefont{J.-C.} \bibnamefont{Tisserand}},
	\bibinfo{author}{\bibfnamefont{M.}~\bibnamefont{Creyssels}},
	\bibinfo{author}{\bibfnamefont{B.}~\bibnamefont{Castaing}}, \bibnamefont{and}
	\bibinfo{author}{\bibfnamefont{F.}~\bibnamefont{Chilla}},
	\bibinfo{journal}{{Phys. Fluids}} \textbf{\bibinfo{volume}{{26}}},
	\bibinfo{pages}{015112} (\bibinfo{year}{{2014}}).
	
	\bibitem[{\citenamefont{Wagner and Shishkina}({2015})}]{wag15}
	\bibinfo{author}{\bibfnamefont{S.}~\bibnamefont{Wagner}} \bibnamefont{and}
	\bibinfo{author}{\bibfnamefont{O.}~\bibnamefont{Shishkina}},
	\bibinfo{journal}{{J. Fluid Mech.}} \textbf{\bibinfo{volume}{{763}}},
	\bibinfo{pages}{109} (\bibinfo{year}{{2015}}).
	
	\bibitem[{\citenamefont{Toppaladoddi et~al.}(2017)\citenamefont{Toppaladoddi,
			Succi, and Wettlaufer}}]{top17}
	\bibinfo{author}{\bibfnamefont{S.}~\bibnamefont{Toppaladoddi}},
	\bibinfo{author}{\bibfnamefont{S.}~\bibnamefont{Succi}}, \bibnamefont{and}
	\bibinfo{author}{\bibfnamefont{J.~S.} \bibnamefont{Wettlaufer}},
	\bibinfo{journal}{Phys. Rev. Lett.} \textbf{\bibinfo{volume}{118}},
	\bibinfo{pages}{074503} (\bibinfo{year}{2017}).
	
	\bibitem[{\citenamefont{Reimann}(2002)}]{rei02}
	\bibinfo{author}{\bibfnamefont{P.}~\bibnamefont{Reimann}},
	\bibinfo{journal}{Phys. Rep.} \textbf{\bibinfo{volume}{361}},
	\bibinfo{pages}{57 } (\bibinfo{year}{2002}), ISSN \bibinfo{issn}{0370-1573}.
	
	\bibitem[{\citenamefont{Prakash et~al.}(2008)\citenamefont{Prakash, Qu\'er\'e,
			and Bush}}]{pra08}
	\bibinfo{author}{\bibfnamefont{M.}~\bibnamefont{Prakash}},
	\bibinfo{author}{\bibfnamefont{D.}~\bibnamefont{Qu\'er\'e}},
	\bibnamefont{and} \bibinfo{author}{\bibfnamefont{J.~W.~M.}
		\bibnamefont{Bush}}, \bibinfo{journal}{Science}
	\textbf{\bibinfo{volume}{320}}, \bibinfo{pages}{931} (\bibinfo{year}{2008}).
	
	\bibitem[{\citenamefont{Eshuis et~al.}(2010)\citenamefont{Eshuis, van~der
			Weele, Lohse, and van~der Meer}}]{esh10}
	\bibinfo{author}{\bibfnamefont{P.}~\bibnamefont{Eshuis}},
	\bibinfo{author}{\bibfnamefont{K.}~\bibnamefont{van~der Weele}},
	\bibinfo{author}{\bibfnamefont{D.}~\bibnamefont{Lohse}}, \bibnamefont{and}
	\bibinfo{author}{\bibfnamefont{D.}~\bibnamefont{van~der Meer}},
	\bibinfo{journal}{Phys. Rev. Lett.} \textbf{\bibinfo{volume}{104}},
	\bibinfo{pages}{248001} (\bibinfo{year}{2010}).
	
	\bibitem[{\citenamefont{Linke et~al.}(2006)\citenamefont{Linke, Alem\'an,
			Melling, Taormina, Francis, Dow-Hygelund, Narayanan, Taylor, and
			Stout}}]{lin06}
	\bibinfo{author}{\bibfnamefont{H.}~\bibnamefont{Linke}},
	\bibinfo{author}{\bibfnamefont{B.~J.} \bibnamefont{Alem\'an}},
	\bibinfo{author}{\bibfnamefont{L.~D.} \bibnamefont{Melling}},
	\bibinfo{author}{\bibfnamefont{M.~J.} \bibnamefont{Taormina}},
	\bibinfo{author}{\bibfnamefont{M.~J.} \bibnamefont{Francis}},
	\bibinfo{author}{\bibfnamefont{C.~C.} \bibnamefont{Dow-Hygelund}},
	\bibinfo{author}{\bibfnamefont{V.}~\bibnamefont{Narayanan}},
	\bibinfo{author}{\bibfnamefont{R.~P.} \bibnamefont{Taylor}},
	\bibnamefont{and} \bibinfo{author}{\bibfnamefont{A.}~\bibnamefont{Stout}},
	\bibinfo{journal}{Phys. Rev. Lett.} \textbf{\bibinfo{volume}{96}},
	\bibinfo{pages}{154502} (\bibinfo{year}{2006}).
	
	\bibitem[{\citenamefont{Lagubeau et~al.}(2011)\citenamefont{Lagubeau, Merrer,
			Clanet, and Qu\'er\'e}}]{lag11}
	\bibinfo{author}{\bibfnamefont{G.}~\bibnamefont{Lagubeau}},
	\bibinfo{author}{\bibfnamefont{M.~L.} \bibnamefont{Merrer}},
	\bibinfo{author}{\bibfnamefont{C.}~\bibnamefont{Clanet}}, \bibnamefont{and}
	\bibinfo{author}{\bibfnamefont{D.}~\bibnamefont{Qu\'er\'e}},
	\bibinfo{journal}{Nature Phys.} \textbf{\bibinfo{volume}{7}},
	\bibinfo{pages}{395} (\bibinfo{year}{2011}).
	
	\bibitem[{\citenamefont{Mar{\'\i}n et~al.}(2012)\citenamefont{Mar{\'\i}n,
			Arnaldo~del Cerro, R{\"o}mer, Pathiraj, Huis~in't Veld, and
			Lohse}}]{marin2012capillary}
	\bibinfo{author}{\bibfnamefont{{\'A}.~G.} \bibnamefont{Mar{\'\i}n}},
	\bibinfo{author}{\bibfnamefont{D.}~\bibnamefont{Arnaldo~del Cerro}},
	\bibinfo{author}{\bibfnamefont{G.~R.} \bibnamefont{R{\"o}mer}},
	\bibinfo{author}{\bibfnamefont{B.}~\bibnamefont{Pathiraj}},
	\bibinfo{author}{\bibfnamefont{A.}~\bibnamefont{Huis~in't Veld}},
	\bibnamefont{and} \bibinfo{author}{\bibfnamefont{D.}~\bibnamefont{Lohse}},
	\bibinfo{journal}{Phys. Fluids} \textbf{\bibinfo{volume}{24}},
	\bibinfo{pages}{122001} (\bibinfo{year}{2012}).
	
	\bibitem[{\citenamefont{H\"anggi and Marchesoni}(2009)}]{han09}
	\bibinfo{author}{\bibfnamefont{P.}~\bibnamefont{H\"anggi}} \bibnamefont{and}
	\bibinfo{author}{\bibfnamefont{F.}~\bibnamefont{Marchesoni}},
	\bibinfo{journal}{Rev. Mod. Phys.} \textbf{\bibinfo{volume}{81}},
	\bibinfo{pages}{387} (\bibinfo{year}{2009}).
	
	\bibitem[{\citenamefont{Ahlers et~al.}(2009)\citenamefont{Ahlers, Grossmann,
			and Lohse}}]{ahl09}
	\bibinfo{author}{\bibfnamefont{G.}~\bibnamefont{Ahlers}},
	\bibinfo{author}{\bibfnamefont{S.}~\bibnamefont{Grossmann}},
	\bibnamefont{and} \bibinfo{author}{\bibfnamefont{D.}~\bibnamefont{Lohse}},
	\bibinfo{journal}{Rev. Mod. Phys.} \textbf{\bibinfo{volume}{81}},
	\bibinfo{pages}{503} (\bibinfo{year}{2009}).
	
	\bibitem[{\citenamefont{Lohse and Xia}(2010)}]{loh10}
	\bibinfo{author}{\bibfnamefont{D.}~\bibnamefont{Lohse}} \bibnamefont{and}
	\bibinfo{author}{\bibfnamefont{K.-Q.} \bibnamefont{Xia}},
	\bibinfo{journal}{Ann. Rev. Fluid Mech.} \textbf{\bibinfo{volume}{42}},
	\bibinfo{pages}{335} (\bibinfo{year}{2010}).
	
	\bibitem[{\citenamefont{van~der Poel
			et~al.}(2015{\natexlab{a}})\citenamefont{van~der Poel, Ostilla-M\'onico,
			Donners, and Verzicco}}]{poe15cf}
	\bibinfo{author}{\bibfnamefont{E.~P.} \bibnamefont{van~der Poel}},
	\bibinfo{author}{\bibfnamefont{R.}~\bibnamefont{Ostilla-M\'onico}},
	\bibinfo{author}{\bibfnamefont{J.}~\bibnamefont{Donners}}, \bibnamefont{and}
	\bibinfo{author}{\bibfnamefont{R.}~\bibnamefont{Verzicco}},
	\bibinfo{journal}{Computers \& Fluids} \textbf{\bibinfo{volume}{116}},
	\bibinfo{pages}{10} (\bibinfo{year}{2015}{\natexlab{a}}).
	
	\bibitem[{\citenamefont{Zhu et~al.}(2017{\natexlab{a}})\citenamefont{Zhu,
			Phillips, Spandan, Donners, Ruetsch, Romero, Ostilla-M{\'o}nico, Yang, Lohse,
			Verzicco et~al.}}]{zhu17afid}
	\bibinfo{author}{\bibfnamefont{X.}~\bibnamefont{Zhu}},
	\bibinfo{author}{\bibfnamefont{E.}~\bibnamefont{Phillips}},
	\bibinfo{author}{\bibfnamefont{V.}~\bibnamefont{Spandan}},
	\bibinfo{author}{\bibfnamefont{J.}~\bibnamefont{Donners}},
	\bibinfo{author}{\bibfnamefont{G.}~\bibnamefont{Ruetsch}},
	\bibinfo{author}{\bibfnamefont{J.}~\bibnamefont{Romero}},
	\bibinfo{author}{\bibfnamefont{R.}~\bibnamefont{Ostilla-M{\'o}nico}},
	\bibinfo{author}{\bibfnamefont{Y.}~\bibnamefont{Yang}},
	\bibinfo{author}{\bibfnamefont{D.}~\bibnamefont{Lohse}},
	\bibinfo{author}{\bibfnamefont{R.}~\bibnamefont{Verzicco}},
	\bibnamefont{et~al.}, \bibinfo{journal}{arXiv:1705.01423}
	(\bibinfo{year}{2017}{\natexlab{a}}).
	
	\bibitem[{\citenamefont{Fadlun et~al.}(2000)\citenamefont{Fadlun, Verzicco,
			Orlandi, and Mohd-Yusof}}]{fadlun00}
	\bibinfo{author}{\bibfnamefont{E.~A.} \bibnamefont{Fadlun}},
	\bibinfo{author}{\bibfnamefont{R.}~\bibnamefont{Verzicco}},
	\bibinfo{author}{\bibfnamefont{P.}~\bibnamefont{Orlandi}}, \bibnamefont{and}
	\bibinfo{author}{\bibfnamefont{J.}~\bibnamefont{Mohd-Yusof}},
	\bibinfo{journal}{J. Comp. Phys.} \textbf{\bibinfo{volume}{161}},
	\bibinfo{pages}{35} (\bibinfo{year}{2000}).
	
	\bibitem[{\citenamefont{Zhu et~al.}(2017{\natexlab{b}})\citenamefont{Zhu,
			Stevens, Verzicco, and Lohse}}]{zhu17b}
	\bibinfo{author}{\bibfnamefont{X.}~\bibnamefont{Zhu}},
	\bibinfo{author}{\bibfnamefont{R.~J. A.~M.} \bibnamefont{Stevens}},
	\bibinfo{author}{\bibfnamefont{R.}~\bibnamefont{Verzicco}}, \bibnamefont{and}
	\bibinfo{author}{\bibfnamefont{D.}~\bibnamefont{Lohse}},
	\bibinfo{journal}{Phys. Rev. Lett.} \textbf{\bibinfo{volume}{119}},
	\bibinfo{pages}{154501} (\bibinfo{year}{2017}{\natexlab{b}}).
	
	\bibitem[{\citenamefont{{{See the supplemental
					material}}}(2017)}]{supplemental}
	\bibinfo{author}{\bibnamefont{{{See the supplemental material}}}}
	(\bibinfo{year}{2017}).
	
	\bibitem[{\citenamefont{Shishkina et~al.}(2010)\citenamefont{Shishkina,
			Stevens, Grossmann, and Lohse}}]{shi10}
	\bibinfo{author}{\bibfnamefont{O.}~\bibnamefont{Shishkina}},
	\bibinfo{author}{\bibfnamefont{R.~J. A.~M.} \bibnamefont{Stevens}},
	\bibinfo{author}{\bibfnamefont{S.}~\bibnamefont{Grossmann}},
	\bibnamefont{and} \bibinfo{author}{\bibfnamefont{D.}~\bibnamefont{Lohse}},
	\bibinfo{journal}{New J. Phys.} \textbf{\bibinfo{volume}{12}},
	\bibinfo{pages}{075022} (\bibinfo{year}{2010}).
	
	\bibitem[{\citenamefont{Johnston and Doering}(2009)}]{johnston2009comparison}
	\bibinfo{author}{\bibfnamefont{H.}~\bibnamefont{Johnston}} \bibnamefont{and}
	\bibinfo{author}{\bibfnamefont{C.~R.} \bibnamefont{Doering}},
	\bibinfo{journal}{Phys. Rev. Lett.} \textbf{\bibinfo{volume}{102}},
	\bibinfo{pages}{064501} (\bibinfo{year}{2009}).
	
	\bibitem[{\citenamefont{Huang et~al.}(2015{\natexlab{a}})\citenamefont{Huang,
			Wang, Xi, and Xia}}]{huang2015comparative}
	\bibinfo{author}{\bibfnamefont{S.-D.} \bibnamefont{Huang}},
	\bibinfo{author}{\bibfnamefont{F.}~\bibnamefont{Wang}},
	\bibinfo{author}{\bibfnamefont{H.-D.} \bibnamefont{Xi}}, \bibnamefont{and}
	\bibinfo{author}{\bibfnamefont{K.-Q.} \bibnamefont{Xia}},
	\bibinfo{journal}{Phys. Rev. Lett.} \textbf{\bibinfo{volume}{115}},
	\bibinfo{pages}{154502} (\bibinfo{year}{2015}{\natexlab{a}}).
	
	\bibitem[{\citenamefont{Shishkina et~al.}(2016)\citenamefont{Shishkina,
			Grossmann, and Lohse}}]{shi16}
	\bibinfo{author}{\bibfnamefont{O.}~\bibnamefont{Shishkina}},
	\bibinfo{author}{\bibfnamefont{S.}~\bibnamefont{Grossmann}},
	\bibnamefont{and} \bibinfo{author}{\bibfnamefont{D.}~\bibnamefont{Lohse}},
	\bibinfo{journal}{Geophys. Res. Lett.} \textbf{\bibinfo{volume}{43}},
	\bibinfo{pages}{1219} (\bibinfo{year}{2016}).
	
	\bibitem[{\citenamefont{Huang et~al.}(2013)\citenamefont{Huang, Kaczorowski,
			Ni, and Xia}}]{huang13}
	\bibinfo{author}{\bibfnamefont{S.-D.} \bibnamefont{Huang}},
	\bibinfo{author}{\bibfnamefont{M.}~\bibnamefont{Kaczorowski}},
	\bibinfo{author}{\bibfnamefont{R.}~\bibnamefont{Ni}}, \bibnamefont{and}
	\bibinfo{author}{\bibfnamefont{K.-Q.} \bibnamefont{Xia}},
	\bibinfo{journal}{Phys. Rev. Lett.} \textbf{\bibinfo{volume}{111}},
	\bibinfo{pages}{104501} (\bibinfo{year}{2013}).
	
	\bibitem[{\citenamefont{van~der Poel
			et~al.}(2015{\natexlab{b}})\citenamefont{van~der Poel, Verzicco, Grossmann,
			and Lohse}}]{poe15jfm}
	\bibinfo{author}{\bibfnamefont{E.~P.} \bibnamefont{van~der Poel}},
	\bibinfo{author}{\bibfnamefont{R.}~\bibnamefont{Verzicco}},
	\bibinfo{author}{\bibfnamefont{S.}~\bibnamefont{Grossmann}},
	\bibnamefont{and} \bibinfo{author}{\bibfnamefont{D.}~\bibnamefont{Lohse}},
	\bibinfo{journal}{J. Fluid Mech.} \textbf{\bibinfo{volume}{772}},
	\bibinfo{pages}{5} (\bibinfo{year}{2015}{\natexlab{b}}).
	
	\bibitem[{\citenamefont{Xi et~al.}(2004)\citenamefont{Xi, Lam, and Xia}}]{xi04}
	\bibinfo{author}{\bibfnamefont{H.-D.} \bibnamefont{Xi}},
	\bibinfo{author}{\bibfnamefont{S.}~\bibnamefont{Lam}}, \bibnamefont{and}
	\bibinfo{author}{\bibfnamefont{K.-Q.} \bibnamefont{Xia}},
	\bibinfo{journal}{J. Fluid Mech.} \textbf{\bibinfo{volume}{503}},
	\bibinfo{pages}{47} (\bibinfo{year}{2004}).
	
	\bibitem[{\citenamefont{Sugiyama et~al.}(2010)\citenamefont{Sugiyama, Ni,
			Stevens, Chan, Zhou, Xi, Sun, Grossmann, Xia, and Lohse}}]{sui10}
	\bibinfo{author}{\bibfnamefont{K.}~\bibnamefont{Sugiyama}},
	\bibinfo{author}{\bibfnamefont{R.}~\bibnamefont{Ni}},
	\bibinfo{author}{\bibfnamefont{R.~J. A.~M.} \bibnamefont{Stevens}},
	\bibinfo{author}{\bibfnamefont{T.~S.} \bibnamefont{Chan}},
	\bibinfo{author}{\bibfnamefont{S.-Q.} \bibnamefont{Zhou}},
	\bibinfo{author}{\bibfnamefont{H.-D.} \bibnamefont{Xi}},
	\bibinfo{author}{\bibfnamefont{C.}~\bibnamefont{Sun}},
	\bibinfo{author}{\bibfnamefont{S.}~\bibnamefont{Grossmann}},
	\bibinfo{author}{\bibfnamefont{K.-Q.} \bibnamefont{Xia}}, \bibnamefont{and}
	\bibinfo{author}{\bibfnamefont{D.}~\bibnamefont{Lohse}},
	\bibinfo{journal}{Phys. Rev. Lett.} \textbf{\bibinfo{volume}{105}},
	\bibinfo{pages}{034503} (\bibinfo{year}{2010}).
	
	\bibitem[{\citenamefont{Ni et~al.}(2015)\citenamefont{Ni, Huang, and
			Xia}}]{ni15}
	\bibinfo{author}{\bibfnamefont{R.}~\bibnamefont{Ni}},
	\bibinfo{author}{\bibfnamefont{S.-D.} \bibnamefont{Huang}}, \bibnamefont{and}
	\bibinfo{author}{\bibfnamefont{K.-Q.} \bibnamefont{Xia}},
	\bibinfo{journal}{J. Fluid Mech.} \textbf{\bibinfo{volume}{778}},
	\bibinfo{pages}{R5} (\bibinfo{year}{2015}).
	
	\bibitem[{\citenamefont{Huang et~al.}(2015{\natexlab{b}})\citenamefont{Huang,
			Wang, Xi, and Xia}}]{hua15}
	\bibinfo{author}{\bibfnamefont{S.-D.} \bibnamefont{Huang}},
	\bibinfo{author}{\bibfnamefont{F.}~\bibnamefont{Wang}},
	\bibinfo{author}{\bibfnamefont{H.-D.} \bibnamefont{Xi}}, \bibnamefont{and}
	\bibinfo{author}{\bibfnamefont{K.-Q.} \bibnamefont{Xia}},
	\bibinfo{journal}{Phys. Rev. Lett.} \textbf{\bibinfo{volume}{115}},
	\bibinfo{pages}{154502} (\bibinfo{year}{2015}{\natexlab{b}}).
	
	\bibitem[{\citenamefont{Zhu et~al.}(2017{\natexlab{c}})\citenamefont{Zhu,
			Verzicco, and Lohse}}]{zhu17}
	\bibinfo{author}{\bibfnamefont{X.}~\bibnamefont{Zhu}},
	\bibinfo{author}{\bibfnamefont{R.}~\bibnamefont{Verzicco}}, \bibnamefont{and}
	\bibinfo{author}{\bibfnamefont{D.}~\bibnamefont{Lohse}}, \bibinfo{journal}{J.
		Fluid Mech.} \textbf{\bibinfo{volume}{812}}, \bibinfo{pages}{279}
	(\bibinfo{year}{2017}{\natexlab{c}}).
	
\end{thebibliography}

\end{document}